\documentclass[11pt]{amsart}
\usepackage{geometry}                % See geometry.pdf to learn the layout options. There are lots.
\usepackage{graphicx}
\usepackage{amssymb}
\usepackage{epstopdf}
\usepackage{etoolbox}
\usepackage{doc}
\begin{document}

\title[Unjamming and nematic flocks during angiogenesis]{Unjamming and nematic flocks in endothelial monolayers during angiogenesis : theoretical and experimental analysis}

\author{Horacio Lopez-Menendez *}{
\thanks{Institut Jacques Monod (IJM), CNRS UMR 7592 et Universit\'e Paris Diderot, 75013 Paris, France}
%\thanks{Department of Mechanics, University of Zaragoza, 50018, Spain}
\thanks{* To whom correspondence should be addressed to Horacio Lopez-Menendez E-mail: horacio.lopez.menendez$@$gmail.com}} 
\author{Joseph D'Alessandro}
%\samethanks{1}}

%\author[1,2,*]{Horacio L\'opez-Men\'endez}}
%
%\author[2]{Jos\'e F\'elix Rodr\'iguez}
%
%
%\thanks[1]{Cell Adhesion and Mechanics, Institut Jacques Monod (IJM), CNRS UMR 7592 \& Universit\`e Paris Diderot, Paris, France}
%\thanks[2]{LaBS, Department of Chemistry, Materials and Chemical Engineering "Giulio Natta", Politecnico di Milano, Piazza Leonardo da Vinci 32, 20133 Milano, Italy}   
%\thanks[*]{Corresponding author: HL-M horacio.lopez.menendez@gmail.com}

\maketitle
%\setlength{\parskip}{2ex plus 0.5ex minus 0.2ex}
%\setlength{\columnsep}{0.7cm}

%\twocolumn[
%\begin{@twocolumnfalse}
%\linenumbers

\begin{abstract}
Angiogenesis is the complex process by which new blood vessels develop from an existing vasculature in order to supply nutrients and/or metabolites to tissues, playing a fundamental role in many physiological and pathological conditions such as embryogenesis and tissue repair or tumour growth. Here we analysed the \textit{in-vitro} angiogenic process from the perspective of the monolayer to understand the role of the interaction between the surrounding endothelial monolayer, the sprouting and the mechanics. First we noticed that the VEGF (Vascular Endothelial Growth Factor) promotes a jamming/unjamming transition that allows the fluidisation of the monolayer by measuring the shape index factor. Next, we measured the density field over the monolayer and realised that the flow of cells manifests strong similarities with the evacuation process where the flock of cells can flow or stuck, defining a convergent channel. Based on these novel observations we propose a mathematical model to describe the effects of unjamming and reorientation of a flock of cells which flow from the monolayer towards the early capillary structure. This model is developed into the framework of the continuum mechanics in which we consider the endothelial monolayer as an active biopolymer film where the nematic order emerges in the unjammed phase promoted by the VEGF activation. To test the proposed ideas we implement the developed coupled equations into a finite element code and describe, with a simplified geometry, the flow effect and cells  orientation from the monolayer to the capillary. In this work, we propose an interpretation, from the top of the endothelial monolayer, based on experimental observations and theoretical arguments to think the angiogenesis.
\end{abstract}

\newpage

\section*{Introduction}

Sprouting angiogenesis is the initiation of micro-capillary growth and it plays an essential role in development reproduction and repair, and is also a prominent feature in a variety of diseases \cite{Folkman1995, carmeliet2000, Ingber2002}. Since the development of the first in-vitro angiogenesis model , many efforts have been made to find the better way to mimic in-vivo situations. It motivated several attempts from the point of view of tissue engineering, where the development of vascular networks represents a bottleneck in its advances to develop organs. Despite this, the growth and expansion of capillary structures in-vivo is a highly dynamic process, showing growth and decrease as a reaction to different types of stimulus. This was observed in physiological conditions during brain development whose electrical activity is associated with the level of vascularisation. In pathological conditions, one of the most sophisticated processes for angiogenesis induction takes place during the tumours growth. Folkman  showed that the tumour induces angiogenesis by the secretion of vascular endothelial growth factors (VEGF) over the adjacent blood vessels, which helps to feed the tumour providing nutrients that let it grow. Therefore, is an extraordinary motivation to figure out the underlying mechanisms that connect different kinds of stimuli - chemical, mechanical, or electrical - with the micromechanical transformations of the endothelia, which allow angiogenesis. In addition, the development of new artificial organs by tissue engineering is limited by the production of blood vessels, needed for supplying oxygen to the cells \cite{laurent2017}. This complex problem has been addressed from the perspective of the genes that play a role in the epithelial junction remodelling. The proteins produced by the genes influence the physical properties of the cell by triggering specific cellular rearrangements. 

In order to gain insight on the development of the  angiogenic process, several mathematical models - mainly continuous and discrete - have been proposed. In particular for wound healing and tumor angiogenesis where angiogenic factors are secreted by the tumor and the extracellular matrix. In general, it can be found continuous and discrete models. The continuous models can describe relevant aspects of angiogenesis at a macroscopic scale, such as average sprout density, vessel growth rates and network expansion~\cite{liotta1977,balding1985, orme1996, holmes2000,mcdougall2002}. In this sense, some recent works provide strong insights on the mechano-chemical interaction~\cite{Checa2009, Mantzaris2004, Das2010,valero2013}, or by chemo-mechanical instabilities \cite{giverso2016}. Better descriptions of the vessel networks can be reproduced with discrete models that account at the single cell scale~\cite{markus1999}. However these models incorporate many phenomenological rules in order to gain accuracy in their prediction.  Moreover, a proper mechanical description of the interactions occurring between the endothelial cells, the vessel structure and the extracellular matrix needs to be incorporated.

Nevertheless, one aspect to be considered that remains elusive, is given by the mechanics of the interaction between the endothelial monolayer, the sprout elongation and the growth \cite{santos2015}. On the one side, the leading cell in the sprouting uses the filopodia to attach to the matrix and exerts force pulling the cells into the endothelial monolayer. On the other side, the cells at the endothelial monolayers proliferate and flow.  Previous works demonstrate that once an endothelial monolayer is activated by an angiogenic factor as VEGF, cells undergo division through an EGF-independent process and possibly push the tip cell forward \cite{Semino2006}. 

In this work we analyse the formation of angiogenic-like structures from the perspective of the endothelial monolayer. We consider that as more cells divide at the monolayer it stores internal stress, then the action of VEGF promotes the change in the tissue from a jammed state towards a fluid-like or unjammed state and the formation of the sprouting. Then, the emergence of sprouting allows mechanical relaxation, which defines a strain field. Therefore, the combination among proliferation and relaxation can be virtuously combined to contribute with the elongation of the capillary.  To do so we reanalyse the images obtained by in vitro experiments with monolayers with endothelial cells under the action of the VEGF performed by Hernandez et al and Semino et al \cite{Semino2006,Herna2009}. In the first part of this work we analyse the endothelial monolayers from the top before and after the application of the VEGF. After that, we develop a theoretical and computational model to describe the observed and interpreted effect over the monolayer. Then in the final part we discuss the results and put them in perspective.

\section*{Experiment analysis}

By analysing the endothelial monolayer from the top view, before and after the application of the VEGF, we noted that it manifests a relevant reorganisation of the cellular packing. Previous works described that the mechanical properties of the tissue can be described in two types: (i) jammed or solid-like structure, if the packing defines a honeycomb-like arrangement. (ii) unjammed or fluid-like if the packing defines an arrangement with squares and triangles with different sizes \cite{farhadifar2007}. The fluidisation of the endothelial monolayer seems to define relevant conditions for the angiogenic process. To clarify the role of the VEGF over the unjamming effect induced in the endothelial monolayer, we followed the same approach as proposed by Park et al \cite{park2015}.  In that work they used the \textit{shape index factor} to study the jamming / unjamming transition in airway epithelium.  The shape index factor comes from the analysis of the vertex model of the cellular monolayer \cite{farhadifar2007,bi2014}. It represents the projection of each cell in two-dimensions by an irregular curved polygon with a shape index, $p=P/\sqrt{A}$, where $P$ and $A$ are the cell perimeter and the projected area, respectively. Here, the mechanical energy of every cell, has three contributions. The first one depends on the bulk elasticity and is proportional to the stretch. The second one, is related to the cortex contractility and depends on the changes in the perimeter. Lastly, an energy term to describe the interfacial cell-cell contact, which defines a net line tension \cite{bi2014}.  The competition between this three terms reaches a preferred cell shape, characterised by its preferred perimeter, $p_0$. 
As the cells exchange neighbours, the structural reorganisation demands alterations in cell shapes. The perturbations in energy to change the cell shapes, represents an energy barrier defined as $\delta\epsilon$.  Depending on the average polygonal shape, the tissue reaches different mechanical properties, more jammed (solid-like) or unjammed (fluid-like). Analysing this model, Manning et al. found a scaling collapse in which the average energy barrier to achieve the cell shape $\delta\epsilon$, exhibits a transition, as a function of the parameter $p\approx 3.81$ which is a critical shape index derived from the analysis of critical scaling \cite{bi2014}.

%%%%%%%%%%%%%%%%%%%%%%%%%%%%%%%

% We demostrate that the VEGF fluidise the monolayer and promote the jamming / unjamming transition

\begin{figure}[h]
\centering
\includegraphics[width=13cm]{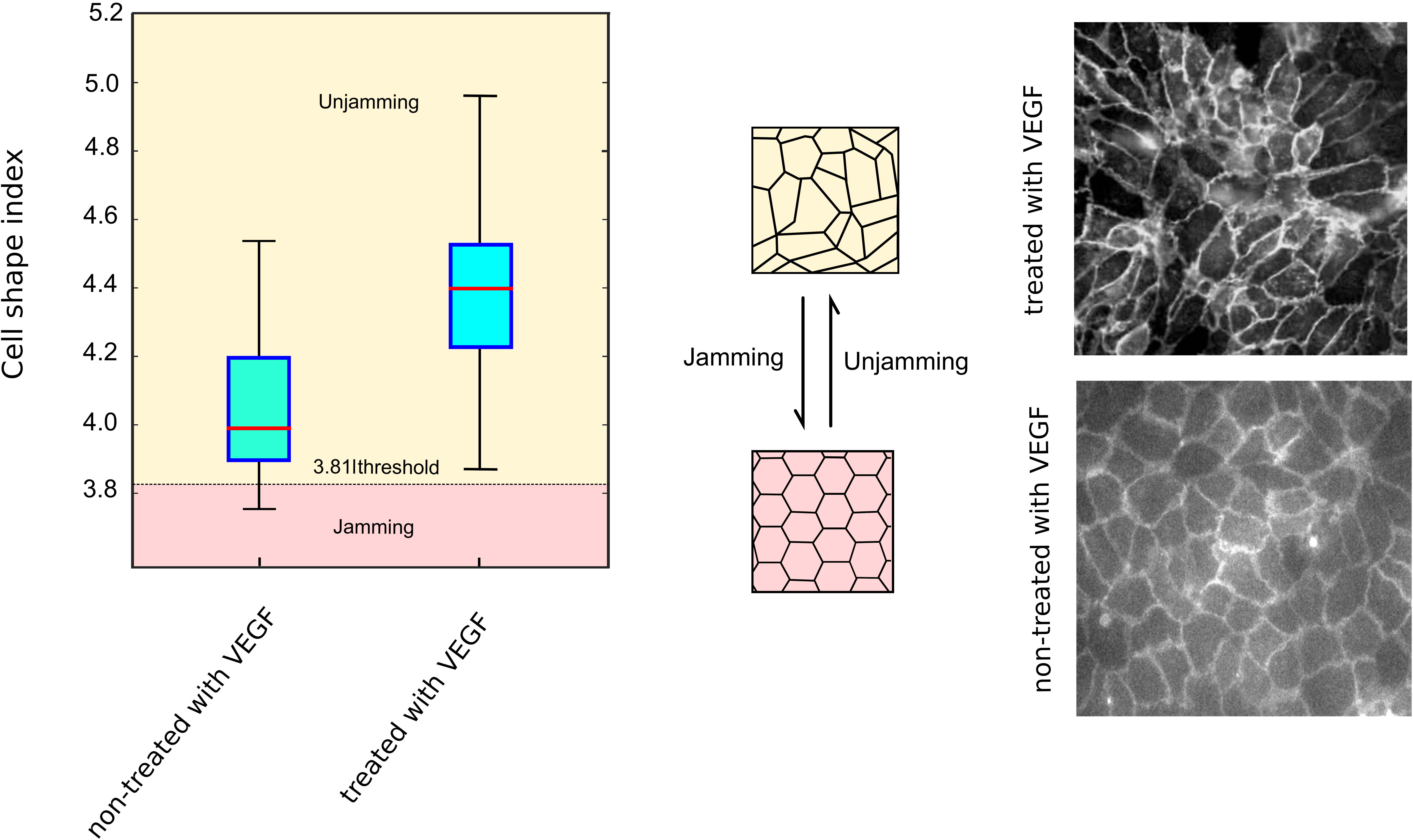}
\caption{ The average shape index factor $p$ is evaluated in the monolayer of endothelial cells with and without the VEGF. It can be observed that in the monolayer non treated with the VEGF, the shape index is more near to the threshold value $(p=3.81)$ to define the transition of Jamming to Unjamming. Nevertheless, the endothelial monolayer treated with the VEGF shows a clear increment on the shape index factor defining a change towards the unjamming and the fluid behaviour. On the right are shown the alterations in the cellular packing due to the treatment with the angiogenic growth factor. The figure non treated manifests honeycomb-like packing, meanwhile the VEGF promotes the fluidisation of the packing. $n \approx 250$, in two different experiment.\label{jamm}}
\end{figure}

Then, in order to test the jamming / unjamming transition of the endothelial monolayer with the action of the VEGF we measured the average shape index $\bar{p}$ over endothelial monolayers with and without the VEGF treatment. To do so we measured the area and the perimeter needed to calculate the shape index factor defining manually the contour of the cells and we calculated its area and perimeter using FIJI image analysis tool. In the Figure 1, it can be observed that for the monolayer without treatment, the shape index is approximately $\bar{p}\approx4$, quite near of the $\bar{p}\approx 3.81$ that represents the jamming transition. Interestingly, for the monolayer treated with the VEGF the value increases until $\bar{p}\approx 4.4$ showing a tendency to raise the level of unjamming or fluidisation of the tissue. The Figure 1, on the right, shows the alterations in the geometrical packing with and without the treatment with the VEGF. Therefore, as a whole, the use of the shape index confirms the plausibility of the jamming / unjamming transition induced by the VEGF. 

%\subsection{Evacuation Dynamics AGREGAR DENSITY FIELD}
Next, we realised that the fluidisation of the monolayer modifies the cells densities distribution. In order to quantify this effect we created a density map for the endothelial monolayer after the application of the VEGF, see Appendix 2 for details. Analysing the density landscape we observed a huge heterogeneity with clusters of cells spanning from more than $\approx 4 [cells/ 1000 \mu m^2]$ (green arrow Figure 2.d) to less than $\approx 1 [cells/1000\mu m ^2]$ (white arrow Figure 2.d).

Then, we reasoned that the change in density is associated with the perturbation introduced by the sprouting. The cells tend to flow towards the sprouting which reduces the density in areas far from it and increases over the surrounding region. This observation shows an unexpected similarity with the "Evacuation process" where a flock of cells flows traversing a small region as is observed for instance when animals, people, particles have to pass through a small gate. In the evacuation processes two effects are well observed. At one side, the formation of a nematic field as a consequence of the queue where the individuals (cells) point in the direction of the gate (sprouting in our case). On the other side, the multiple convergence of individuals disturbs each other, stucks and frustrates the passage creating a bottle neck (density increase) around the traversing gate.

\begin{figure}
\centering
\includegraphics[width=13cm]{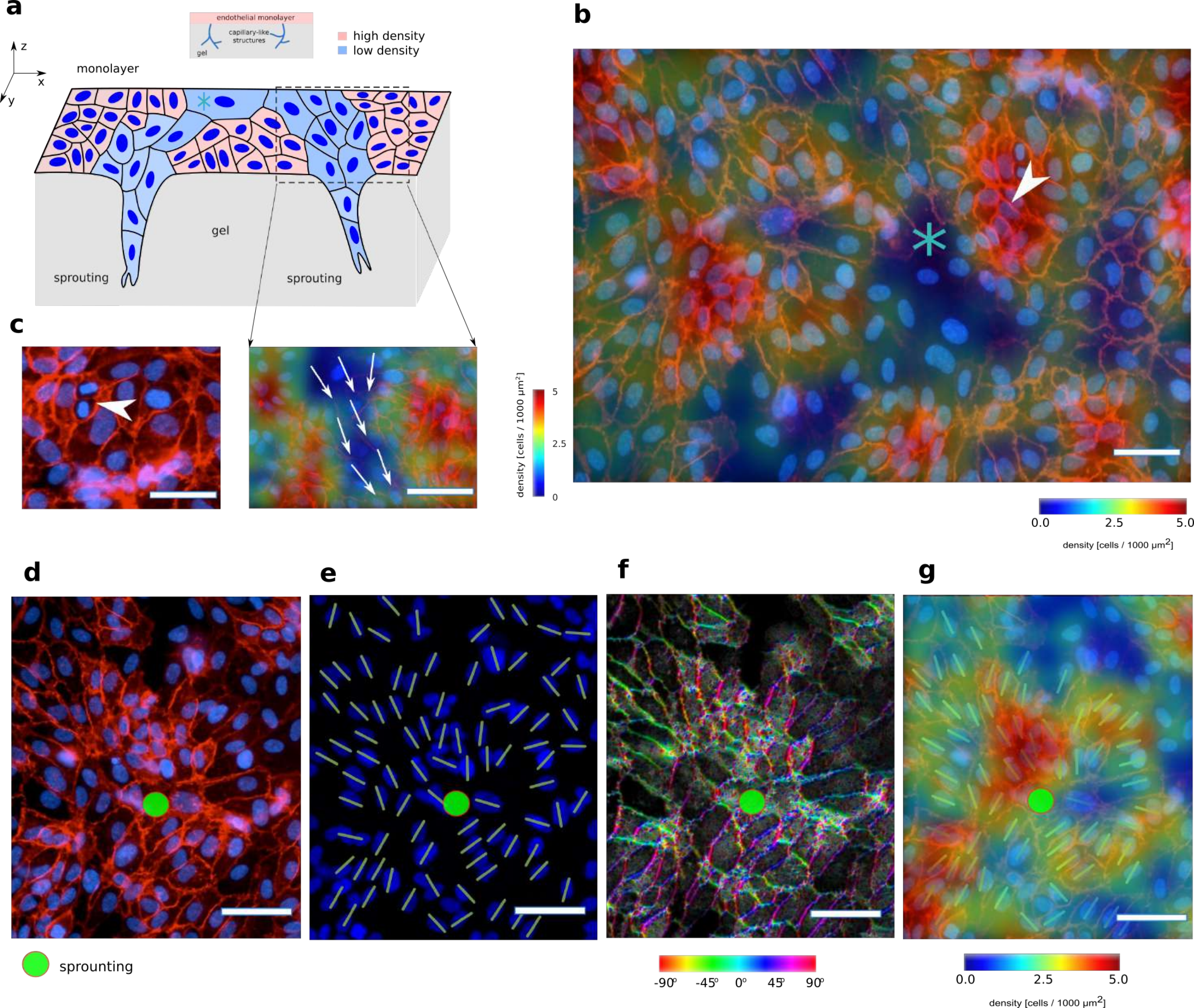}
\caption{ \textbf{a.} The figure describes a portion of the experimental setup, where the endothelial monolayer is cultivated over a gel. Due to the action of the VEGF the interaction between the monolayer and the capillary-like structures promotes alterations over the cell densities distribution. The dotted box represents the region on the monolayer defining a sort of convergent channel. \textbf{b.} Cells density distribution landscape: The figure shows the organisation of the cell under the action of the VEGF. It can be observed the formations of clusters of high, medium and low densities. The green asterisk suggests that the existence of very low density regions are due to the flow of cells toward the sproutin regions. The white arrow points a cluster with very high density. \textbf{c.} The image shows the coupling effect between the cell division on the monolayers and its contribution with the formation of the capillary-like structure. The white arrow points the regions of the sprouting and the white arrow a cell under division. \textbf{d.} Organisation of the cells in the neighbourhood of the sprouting, in the endothelial monolayer treated with the VEGF. \textbf{e.} Visualisation of the cellular nucleus showing the formation of a nematic field where the nucleus manifests a clear correlation between the nuclear orientation pointing towards the sprouting.  \textbf{f.} Orientation field defined by the angular correlation intensity defined by the cellular structure. Here, we further confirm the alignment of a flock of cells in the direction of the sprouting. To do so we did image analysis by ImageJ with the plugin OrientationJ.  \textbf{g.} Combination of the density map and the nematic orientations.}
%\label{jamm}}
\end{figure}

To better quantify these clues, we studied the orientation of the cells towards the sprouting. Previous works showed that the nucleus can be used as a compass to determine the anisotropic strains over the cytoskeleton, at single cell level \cite{versaevel2012}, or inside a tissue \cite{vedula2014}. When we look the nuclear orientations we clearly observe the definition of a nematic field in the surroundings of the sprouting until a distance of tens of cells, (see Figure 2e). Also, this can be further proved using image analysis with FIJI with the OrientationJ plugin to determine the orientations of the cellular cortex in a region surrounded by the sprouting, where the cellular alignment confirms the formation of the nematic field (Figure 2f).

When we analysed jointly the nematic field with the density map (Figure 2g), we notice that the cells in the density field are less dense which is equivalent to say that they are more stretched. It suggests that the cells are into a strain field among the monolayer and the capillary-like structure, where the strain field let the cells flow and reorient themselves defining the queue. Furthermore, aside the sprouting we find clusters of cells with high density, which we interpret as frustrated cells that are not able to flow (bottle neck). Another feedback that emerges from the reorientation of the cells towards the sprouting, is the fact that the cell under division will growth in the direction of the sprouting. It can be observed in the Figure 2.c. Taking these observations as a whole, we realise that the angiogenic process observed from the perspective of the monolayer can be described as a flow of cells towards the capillary-like structure through the channel created between the frustrated flock, as can be observed in the Figure 2.

%%%%%%%%%%%%%%%%%%%%%%%%%%%%%%%%%
%%% VERIFICAR QUE LAS FIGURAS ESTEN BIEN %%%%%%
%%%%%%%%%%%%%%%%%%%%%%%%%%%%%%%%%

%\subsection{Simplified geometrical model}

Here, we propose a new mathematical model to describe the angiogenic process. The combination between cell proliferation and confluent monolayer increases the internal cell stress and defines a jammed state. Under the action of the VEGF, the endothelial cells manifest a phase transition from the jamming to unjamming and promote the formation of a sprouting. The sprout enhances the relaxation of the trapped internal stress defining a strain field. Under the presence of a strain field, the unjammed cells tend to align in the direction of the principal strain, pointing in the direction of the sprouting. Hereafter, the aligned flock of cells define similarities among the nematic field and the evacuation process. The combination between the stuck and frustrate clusters of cells and the flock of nematic cells, defines a channel where the cells are able to flow, through geometrical constrains, towards the sprouting (Figure 2.b,d). 

\section*{Theoretical and computational model for tissue monolayer}

Here, we determine the material model for the monolayer to provide a physical explanation of the proposed tissue dynamics. Some previous models appeal to analogies with other physical systems as, liquid crystals~\cite{Ranft2010} or glassy transitions~\cite{Trepat2011,Angelini2011}.  We propose a material model into the framework of non-linear continuum mechanics which considers the endothelial monolayer as an active biopolymer film where a nematic order emerges from the VEGF concentration. In the following subsections we explain the detailed ingredients of the mathematical model: (i) an introduction of the main variables used in the formalism of non-linear continuum mechanics. (ii) In order to model the different packing of clusters of cells defining isotropic and nematic orientations, we propose an homogenisation scheme based on a structure tensor. We define an order parameter to characterise the degree of the unjamming phase induced by the VEGF over the endothelial monolayer.  This effect is deduced by a phase field model via a Landau-like free energy function coupled with a non-linear strain energy function. (iii) reorientation model: to describe the reorientation of the nematic order vector we introduce an evolution equation. This expression describes the alignment in the direction of the maximum principal strain. (iv) continuum inelastic model: To describe the homogenized monolayer we introduce a non-linear mechanics model based on the semiflexible worm-like chain model, in which we incorporate the active pre-strain introduced by the phosphorylated myosin. Next, we introduce the inelastic effects that emerge when the flock of cells pass across "the gate" to flow from the monolayer towards the sprouting-like structure. (v) Numerical simulation: To test the proposed ideas, we implemented the coupled equations into a finite element code, in a simplified geometry.

\subsubsection*{Basic results of the continuum mechanics} \label{sec_basic}
Let $\mathcal{B}_0$ be a continuum body defined as a set of points in a certain
assumed reference configuration. Denote by
$\{\boldsymbol{\chi}:\mathcal{B}_0\rightarrow\mathcal{R}^3\}$ the continuously
differentiable, one to one mapping (as well as its inverse
$\boldsymbol{\chi}^{-1}$) which puts into correspondence $\mathcal{B}_0$ with
some region $\mathcal{B}$, the deformed configuration, in the Euclidean space.
This one-to-one mapping $\boldsymbol{\chi}$ transforms a material point
$\textbf{X}\in\mathcal{B}_0$ to a position
$\textbf{x}=\boldsymbol{\chi}(\textbf{X})\in\mathcal{B}$ in the deformed
configuration.

The deformation gradient $\textbf{F}$ is defined as
\begin{equation}
    \textbf{F}:=\frac{\partial\boldsymbol{\chi}(\textbf{X})}{\partial\textbf{X}},
\end{equation}
with $J(\textbf{X})=\det(\textbf{F})>0$ the local volume ratio. It is sometimes useful to consider the multiplicative
split of $\mathbf{F}$
\begin{equation}\label{F_split}
    \textbf{F}=J^{1/3}\textbf{1}\bar{\textbf{F}},
\end{equation}
into dilatational and distortional (isochoric) parts, where $\textbf{1}$ is the
second-order identity tensor. Note that $\det(\bar{\textbf{F}})=1$. From this,
it is now possible to define the right and left Cauchy-Green deformation
tensors, $\textbf{C}$ and $\textbf{b}$ respectively, and their corresponding
isochoric counterparts $\bar{\textbf{C}}$ and $\bar{\textbf{b}}$
\begin{equation}\label{TensoresCyb}
\begin{array}{ll}
\textbf{C}=\textbf{F}^T\textbf{F}=J^{2/3}\bar{\textbf{C}}, &
\quad \bar{\textbf{C}}=\bar{\textbf{F}}^T\bar{\textbf{F}}, \\
\textbf{b}=\textbf{F}\textbf{F}^T=J^{2/3}\bar{\textbf{b}}, & \quad
\bar{\textbf{b}}=\bar{\textbf{F}}\bar{\textbf{F}}^T,
\end{array}
\end{equation}
it being straightforward to show that
\begin{equation}\label{RelacionCCbar}
\frac{\partial J}{\partial \textbf{C}}=\frac{1}{2}J \textbf{C}^{-1}, \qquad \frac{\partial
\overline{\textbf{C}}}{\partial \textbf{C}}=J^{-2/3}\left(\mathbb{I}-\frac{1}{3}
\overline{\textbf{C}}\otimes\overline{\textbf{C}}^{-1}\right),
\end{equation}
For a hyperelastic material, the stress at a point
$\textbf{x}=\boldsymbol{\chi}(\textbf{X})$ is only a function of the
deformation gradient $\textbf{F}$ at that point. A change in stress obeys only
to a change in configuration. In addition, for isothermal and reversible
processes, there exists a scalar function, a strain energy function (SEF) $\Psi$, from which the
hyperelastic constitutive equations at each point $\textbf{X}$ can be derived.
The function $\Psi$ must obey the \emph{Principle of Material Frame
Indifference} which states that constitutive equations must be invariant under
changes of the reference frame
\begin{equation}
\Psi(\textbf{X},\textbf{C})=\Psi(\textbf{X},\textbf{QCQ}^T),\;\;\forall(\textbf{Q},\textbf{C})
\in\mathrm{Q}^+\times\mathrm{S}^+,
\end{equation}
where
\begin{equation}
\begin{array}{c}
\mathrm{S}^+=\{\textbf{C}\in\mathcal{L}(\mathbb{R}^3,\mathbb{R}^3):\textbf{C}=\textbf{C}^T,
\textbf{C}\;\mathrm{positive\;definite}\},\\
\mathrm{Q}^+=\{\textbf{Q}\in\mathcal{L}(\mathbb{R}^3,\mathbb{R}^3):\textbf{Q}^T\textbf{Q}=1\},\\
\end{array}
\end{equation}
and $\mathcal{L}(\mathbb{R}^3,\mathbb{R}^3)$ denotes the vector space of linear
transformations in $\mathbb{R}^3$. For materials with a particular symmetry
group, the dependence of $\Psi$ on the deformation gradient is affected by the
symmetry group itself. For the case of transversely isotropic materials, the
directional dependence of the SEF on the deformation (restriction imposed by
the symmetry) is commonly defined by introducing a vector representing the
material preferred direction. In this regard, let $\mathbf{a}_0$ be unit vector field describing the local fiber direction
in the undeformed configuration. The SEF can now be expressed as an isotropic function of the
right Cauchy-Green deformation tensor and the vector field $\mathbf{a}_0$ as
\begin{equation}
\Psi(\textbf{X},\textbf{C},\textbf{a}_0)=\Psi(\textbf{X},\textbf{QCQ}^T,
\textbf{Q}\textbf{a}_0\otimes\textbf{a}_0\textbf{Q}^T),
\end{equation}
for all $(\textbf{Q},\textbf{C}) \in\mathrm{Q}^+\times\mathrm{S}^+$, and where
$\otimes$ represents the tensor outer product. Further,
Spencer \cite{Spencer1980} showed that the irreducible integrity bases for the 
symmetric second-order tensors $\mathbf{C}$ and $\mathbf{a}_0\otimes\mathbf{a}_0$, correspond to four
invariants:
\begin{equation}\label{invariantes}
\begin{array}{c}
  \begin{array}{ccccc}
  I_1=\mathrm{tr}\textbf{C}, &&
  I_2=\frac{1}{2}[(\mathrm{tr}\textbf{C})^2-
  \mathrm{tr}\textbf{C}^2], && I_3=
  \det\textbf{C}=1, \\
 && I_4=\textbf{a}_0\cdot\textbf{C}\cdot\textbf{a}_0. &&
  \end{array}
\end{array}
\end{equation}
Invariants $I_1$, $I_2$, $I_3$ are standard invariants of the Cauchy-Green deformation
tensor, and are associated with the isotropic material behavior. Invariant
$I_4$, arises from the anisotropy introduced by the fiber. 
With this invariants at hand, a SEF can now be proposed satisfying the
material frame indifference and the material symmetry restrictions
\begin{equation}\label{SEF_anisotropic}
\Psi(\textbf{X},\textbf{C},\textbf{a}_0)=\Psi(\textbf{X},I_1,I_2,
I_3,I_4).
\end{equation}
This relation provides the basis for the development of constitutive models
carried out in the following sections, as well as for deriving the stress and
elasticity tensors used in the Finite Element implementation. As a final
remark, for quasi-incompressible materials, dilatational and deviatoric parts
of the deformation gradient receive a separate numerical treatment because of
the ill-conditioning caused by the dilatational stiffness. In this regard it was proposed a
representation of quasi-incompressible elasticity in which the SEF takes an
uncoupled form in which the dilatational and deviatoric parts are such that
\begin{equation}\label{SEF_anisotropic_uncoup}
\Psi(\textbf{X},\textbf{C},\textbf{a}_0)=U(J)+\bar{\Psi}(\textbf{X},
\bar{I}_1,\bar{I}_2,\bar{I}_4)
\end{equation}
where $\bar{I}_k, k=1,\ldots,4$, are the invariants of the isochoric
Cauchy-Green tensor $\bar{\textbf{C}}$ (note that $\bar{I}_3=1$). In the
developments in the next section, we use a SEF of the form given in Eq.~\ref{SEF_anisotropic_uncoup}.

\subsubsection*{Homogenized description of cells-monolayer }

\begin{figure}[h]
\centering
\includegraphics[width=10cm]{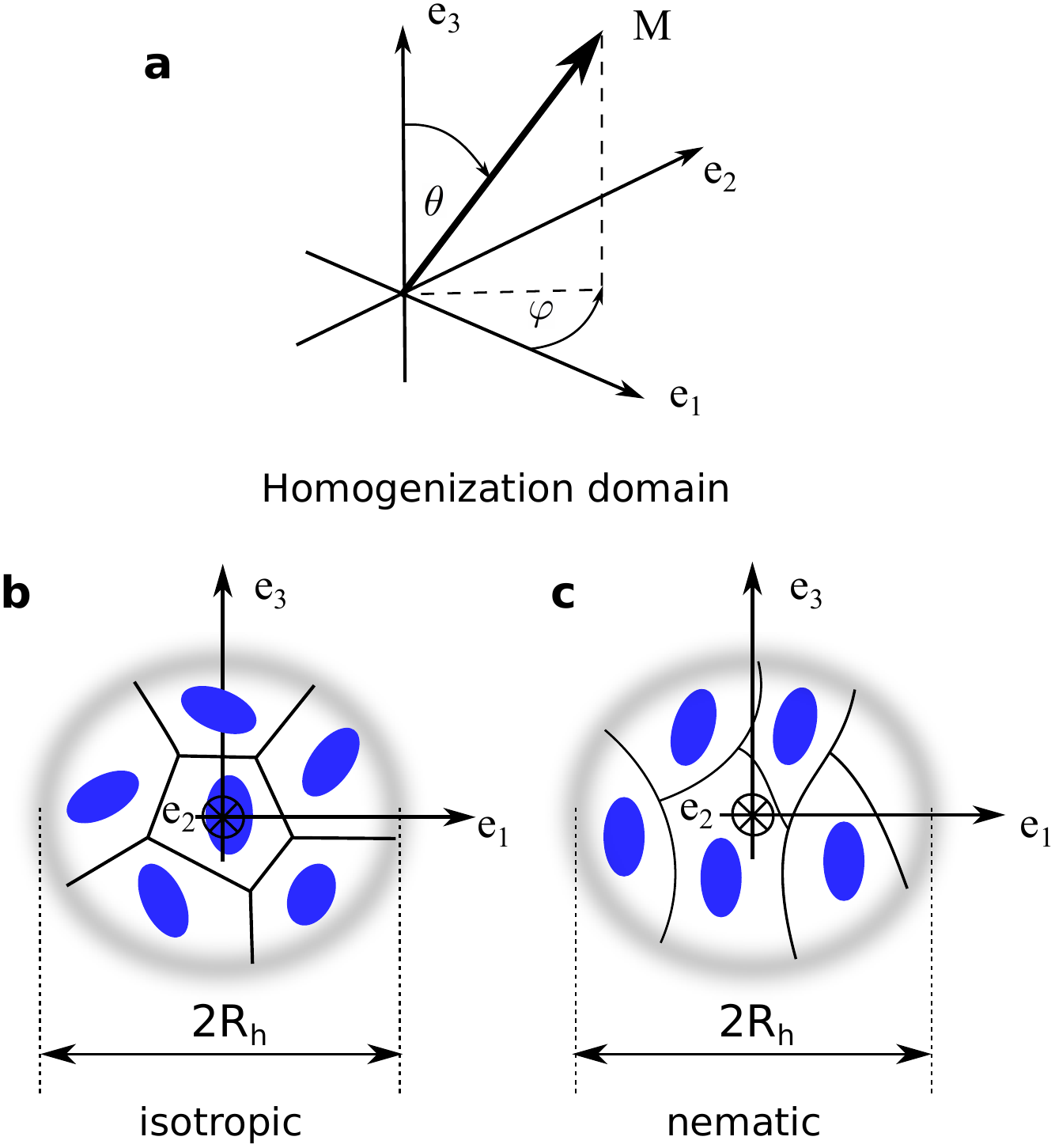}
\caption{a. Cartesian representation of the $\mathbf{M}$ by means of Euler angles. Figures b and c represents the homogenization domain associated with an isotropic and nematic distribution of cell organization respectively.}\label{homog}
\end{figure}

In order to model the different sort of packings into the endothelial monolayer defining isotropic or nematic orientations, from the perspective of the continuum mechanism, we introduce and homogenisation scheme based on a structure tensor, as proposed by Gasser et al \cite{Gasser2006} for biological soft-tissues.  An orientation vector $\mathbf{M}$ with unit length is introduced and characterised through the Euler angles $\theta\in[0,\pi/2]$ and $\varphi\in[0,2\pi]$ (see Figure~\ref{homog}) according to 
\begin{equation}
\mathbf{M}(\theta,\varphi)=\sin\theta \cos\varphi\mathbf{e_{1}}+\sin\theta \sin\varphi\mathbf{e_{2}}+\cos\theta\mathbf{e_{3}}
\end{equation}
where $\{\mathbf{e_{1}},\text{\ensuremath{\mathbf{e_{2}}},}\text{\ensuremath{\mathbf{e_{3}}}}\}$
are the orthogonal basis in a rectangular Cartesian coordinate system.

The distribution of cells in the reference volume is implemented through an orientation density function $\rho(\mathbf{M})$ with respect to the referential orientation unit vector $\mathbf{M}$ and it is normalized over the half unit sphere such that. 
\begin{equation}
\intop_{\varphi=0}^{2\pi}\intop_{\theta=0}^{\pi/2}\rho\left(\theta,\varphi\right)\sin\theta d\theta d\varphi=1
\end{equation}

and the representative homogenisation radius of the micro-sphere $R_h$ is higher than the average cell size, as it is shown in Figure \ref{homog} \cite{Murtada2010a}. The average strain in the homogenisation radius $R_h$ can be expressed as

\begin{equation}
\left< \lambda  \right> =\intop_{\varphi=0}^{2\pi}\intop_{\theta=0}^{\pi/2}\rho\left(\theta,\varphi\right)\mathbf{M}(\theta,\varphi).\mathbf{CM}(\theta,\varphi)\sin\theta d\theta d\varphi=\mbox{\ensuremath{\sqrt{\left(\mathbf{C:H}\right)}}}
\end{equation}
where $\mathbf{H}$ is a symmetric generalised structure tensor representing the nematic order into the micro-sphere $R_h$.

\begin{equation}
\mathbf{H}=\intop_{\varphi=0}^{2\pi}\intop_{\theta=0}^{\pi/2}\rho\left(\theta,\varphi\right)\mathbf{M}(\theta,\varphi)\otimes\mathbf{M}(\theta,\varphi)\sin\theta d\theta d\varphi
\end{equation}
Assuming rotational symmetry about the mean referential vector $\mathbf{a_{0}}$ (it means that, the oriented flock of cells behaves as a transversely isotropic material), and, without loss generality, taking this preferred direction
$\mathbf{a_{0}}$ to be coincident with the cartesian basis vector $\mathbf{e_{3}}$,
the density function, $\rho$, can be described independently of the Euler
angle $\varphi$, i.e $\rho\mathbf{M}(\theta,\varphi)=\rho\left(\theta\right)$. 
This leads, after some manipulations (see Gasser et.al~\cite{Gasser2006}), to
\begin{equation}
\mathbf{H}=\frac{1}{3}\left(1-\Gamma\right)\mathbf{I}+\Gamma\mathbf{a_{0}}\otimes\mathbf{a_{0}}
\end{equation}
where 
\begin{equation}
\Gamma =1-3\pi \int _{ 0 }^{ \pi /2 }{ \rho (\theta )sin^{ 3 }\theta d\theta} 
\end{equation}

Where $\mathbf{I}$ denotes the second order identity tensor. Hence $\mathbf{H}$
depends on a single orientation parameter, associated with the characteristics
of the structure $\Gamma$, which represents the orientation of the flock of cells in a integral sense and describes its degree of anisotropy. As we can see, the structure tensor $\mathbf{H}$ represents a linear mixture between an isotropic and the anisotropy description associated with the organization of the flock of cells. With $\Gamma\in\left[ 0,1 \right]$. As inherent to homogenisation schemes, the information is lost when performing transition from $\rho$ to $\mathbf{H}$. 

%Further, we hypothesise that there exists relationship between the VEGF and the change in the orientability of the cells that facilitates the collective cell migrations in the direction of the sprouting. This will be introduced mathematically into the model as a change in the value of $\Gamma$, following the suggested relation showed in Figure \ref{anisotropy_vegf}.

%Accordingly, the computation of $\rho$ for $\mathbf{H}$ given is not unique but constitutes an essential part of the modelling it self. In this step we will not focus in the appropriate determination of $\Gamma$.

%Further studies will take measurements to correlate the cell orientation with the concentrations of VEGF, this studies will determine the concentration parameter via the von-Mises  distribution and the concentration of VEGF.

The averaged strain over the cluster of cells is defined through the structure tensor $\mathbf{H}$ as a function of the invariants $\bar{I}_1$ and $\bar{I}_4$ as:
\begin{equation}
\left< \lambda\right> =\sqrt{\mathbf{C}:\mathbf{H}}
= \sqrt{\frac{1}{3}(1-\Gamma) \bar{I}_1+\Gamma \bar{I}_4 }
\end{equation}
This expression for the strain will be used, in the followings sections, to evaluate the strain energy function associated within the cell cluster in the finite element formulation.

\subsubsection*{Energy function and order parameter}

Here we propose a free energy function considering one term for the phase transition for the jamming/unjamming based on Landau functional and another for the strain energy function. The Landau theory is a mean field theory without including all the degrees of freedom involved in the statistical model. It is a phenomenological theory written in terms of an order parameter which in our case is defined as $unjamming-level$  $\Gamma=\Gamma(c)$ where $c$ represents the concentrations of the VEGF. From symmetry considerations alone the condition of thermodynamical equilibrium is realised as a minimisation of the free energy. Following, we write down a generic Landau functional of the free energy that couples the strain with the structure tensor as:

\begin{equation}
\Psi( \mathbf{C}, \mathbf{H};\Gamma)=\frac{\alpha}{2}\Gamma^2 + \frac{\beta}{4}\Gamma^4 + \Psi_{nem}( \mathbf{C}, \mathbf{H};\Gamma) + \Psi_{iso}(\mathbf{C})\label{ener}
\end{equation}

Whereas the first two terms describe the energy associated with the order parameter, the third term describes the strain energy function for the nematic structure as a function of the Cauchy strain tensor and the structure tensor. The fourth term represents the isotropic strain energy function to describe the elasticity of the rest of elements with isotropic elastic properties.

Since the equilibrium position (minimum) of $\Psi(\Gamma,c)$ changes at $\alpha=0$, we identify $\alpha=0$ with the critical point $c=c_{cr}$. This allow us to choose $k\hat{c}$ as $\alpha$, where $k$ is a positive constant and $\hat{c}=(c-c_{cr})/c_{cr}$ is the deviation of the concentration ratio from the critical point normalised by $c_{cr}$ which we define as a reduced concentration ratio. Then, the simplest election is $\alpha=k\hat{c}$, for which $\alpha>0$ above the critical point and $\alpha<0$ below. The dependence of $\beta$ with $\hat{c}$ does not affect qualitatively the behaviour of the free energy in the vicinity of the critical point and therefore we take $\beta$ as a constant. Then, minimising the free energy to obtain the equilibrium condition, we derive the expression $\ref{ener}$ with respect to the order parameter $\Gamma$

\begin{equation}
\frac{\partial \Psi}{\partial \Gamma}\approx2\alpha\Gamma+4\beta\Gamma^3=0.
\end{equation}

Thus, the equilibrium value of remodeling, $\Gamma$ is

\begin{equation}
\Gamma\approx\left( \frac{-\alpha}{2\beta}\right)^{1/2}=\left[\frac{k(c-c_{cr})}{2\beta c_{cr}}\right]^{1/2};\ \forall \  (c > c_{cr})
\end{equation}

\begin{figure}[h]
\centering
\includegraphics[width=8cm]{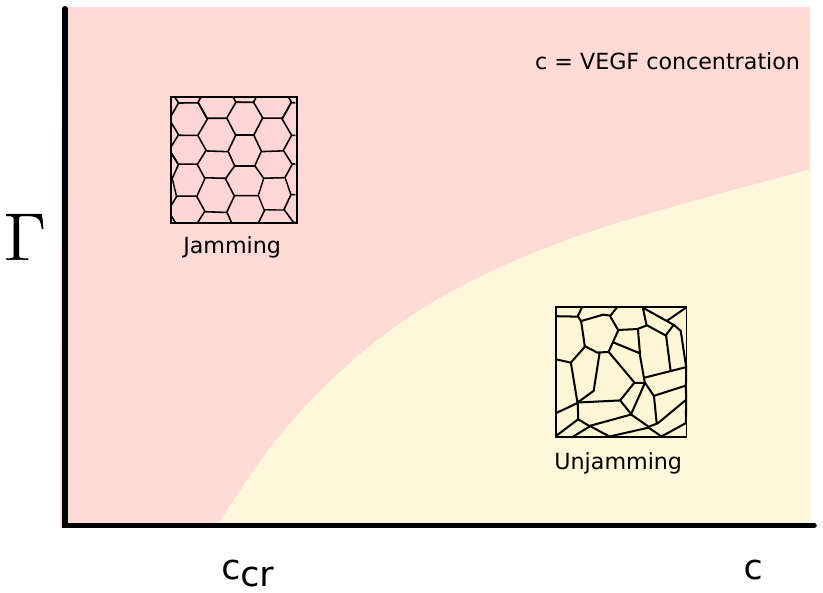}
\caption{Phase transition \label{anisotropy_vegf}}
\end{figure}

Then as much as the concentration of the VEGF increases above the critical concentration-i.e., $c>c_{cr}$ the order parameter, $\Gamma(c)$ that describes the degree of unjamming, increases as a square root of $c-c_{cr}$. It implies that the lost of isotropic and hexagonal-like packing allows the ability of the cell to be orientable. As can be observed in the structure tensor, as much as $\Gamma(c)$ increases the nematic order increases as well and becomes orientable, as is described by the structure tensor $\mathbf{H}=\frac{1}{3}\left(1-\Gamma\right)\mathbf{I}+\Gamma\mathbf{a_{0}}\otimes\mathbf{a_{0}}$. In the following we introduce the evolution law to describe the reorientation of the nematic phase.

\subsubsection*{Reorientation}

Here we propose a way to describe the evolution of the nematic orientation for homogenised flock of cells. 
We consider the case where the relaxation of the orientation tensor is driven mainly by the local anisotropic strain, and we ignore additional effects such as morphogen gradients. Also,  we consider the case where the anisotropy relaxation is faster than cell division. Following the model suggested by Kuhl et al. \cite{Kuhl2005, Kuhl2007}, the vector ${\mathbf{a}_{o}}$ is allowed to gradually align with the eigenvector ${\mathbf{a}}_{\lambda}^{\max}$ corresponding to the maximum eigenvalue, $\lambda^{\max}$, of Cauchy Green tensor $\mathbf{C}$.
\begin{figure}[ht]
\centering
\includegraphics[width=10cm]{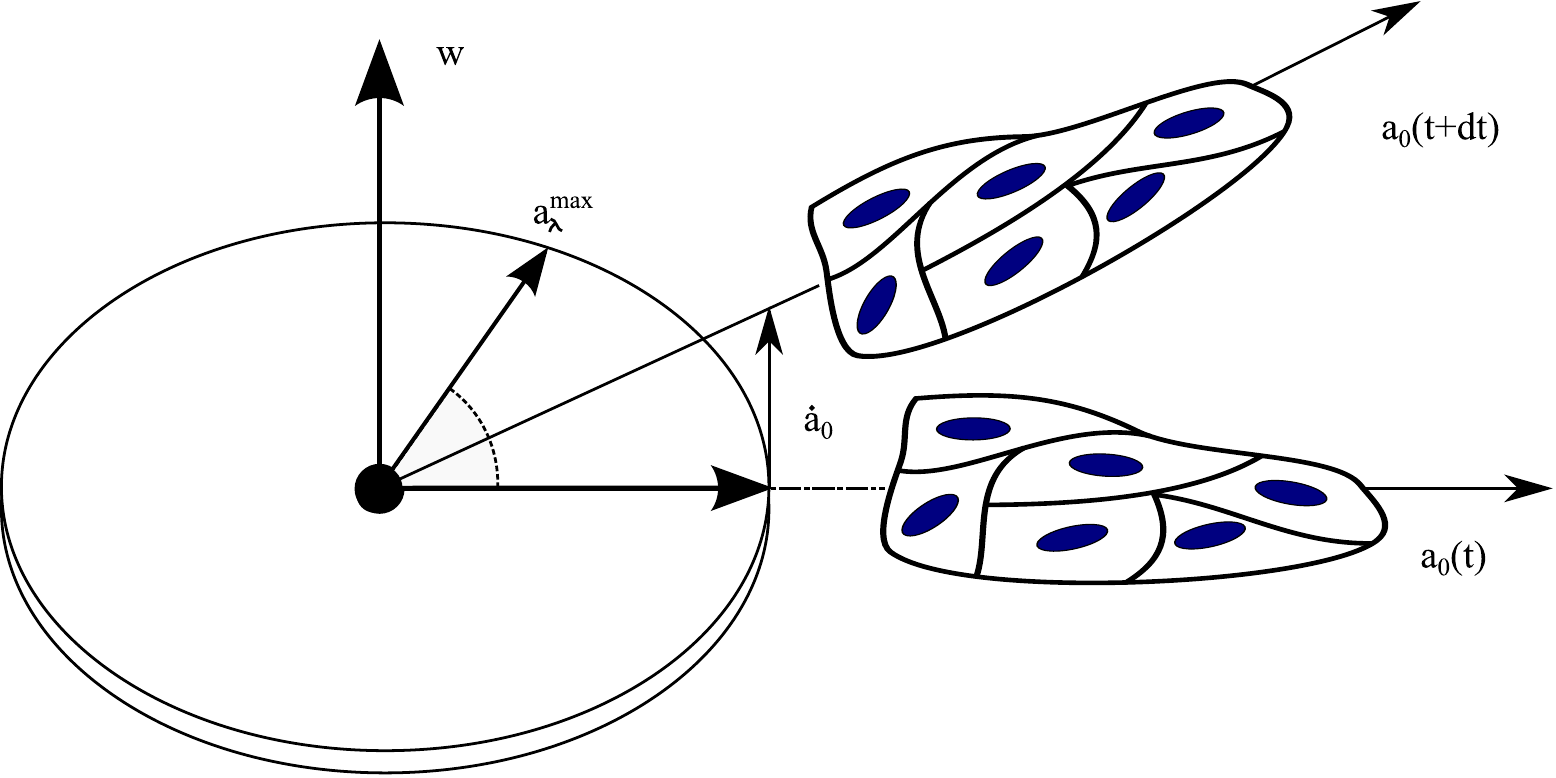}
\caption{Reorientation of the clusters of cells with respect to the principal strains.}
\end{figure}
The joint action between the deformation induced by the sprouting and the monolayer defines a strain field where the flock of cells tends to align in the direction of the principal of strain. 
Then, we introduce the rotational vector as:
\begin{equation}
\omega=\frac{1}{\tau_{\omega}}\left({\mathbf{a}_{0}}\wedge{\mathbf{a}_{\lambda}^{max}}\right)
\end{equation}
where $\omega$ is the scaled vector product of the stress-fiber orientation vector,
${\mathbf{a}_{0}}$, and the maximum principal strain, $\mathbf{a}_{\lambda}^{max}$. Then, the rotation
vector can be expressed as
\begin{equation}
{\mathbf{\omega}}=\omega{\mathbf{a}}^{\omega}
\end{equation}
\begin{equation}
{\mathbf{a}}^\omega=\frac{{\mathbf{a}_{0}}\wedge{\mathbf{a}}_{\lambda}^{max}}{\parallel{\mathbf{a}_{0}}\wedge{\mathbf{a}}_{\lambda}^{max}\parallel}
\end{equation}
\begin{equation}
\omega=\frac{{\parallel{\mathbf{a}_{0}}\wedge{\mathbf{a}}_{\lambda}^{max}\parallel}}{\tau_{\omega}}
\end{equation}
The evolution for the flock of cells orientation can be expressed following the
abstract form 
\begin{equation}
{\mathbf{\dot{a}}_{o}}=D_{t}{\mathbf{a}_{0}}={\mathbf{\omega}a_{0}}=-[\mathbf{e}.\omega].{\mathbf{a}_{0}}
\end{equation}
where ${\mathbf{e}}$ denotes the third-order permutation symbol, and the orthogonality condition $D_{t}{\mathbf{a}_{0}}.{\mathbf{a}_{0}}=0$
is valid throughout the remodeling history. Therefore the previous expression can be rewritten
as:
\begin{equation}
{\mathbf{\dot{a}}_{o}}=\frac{1}{\tau_{\omega}}\left[{\mathbf{a}_{\lambda}^{max}}-
({\mathbf{a}_{\lambda}^{max}}.{\mathbf{a}_{0}}){\mathbf{a}_{0}}\right]
\end{equation}
Then we arrive to an expression for the evolution of the nematic orientation in the direction of the principal strain.

\subsubsection*{Inelastic model for homogenized flock of cells} 

\begin{figure}[ht]
\centering
\includegraphics[width=10cm]{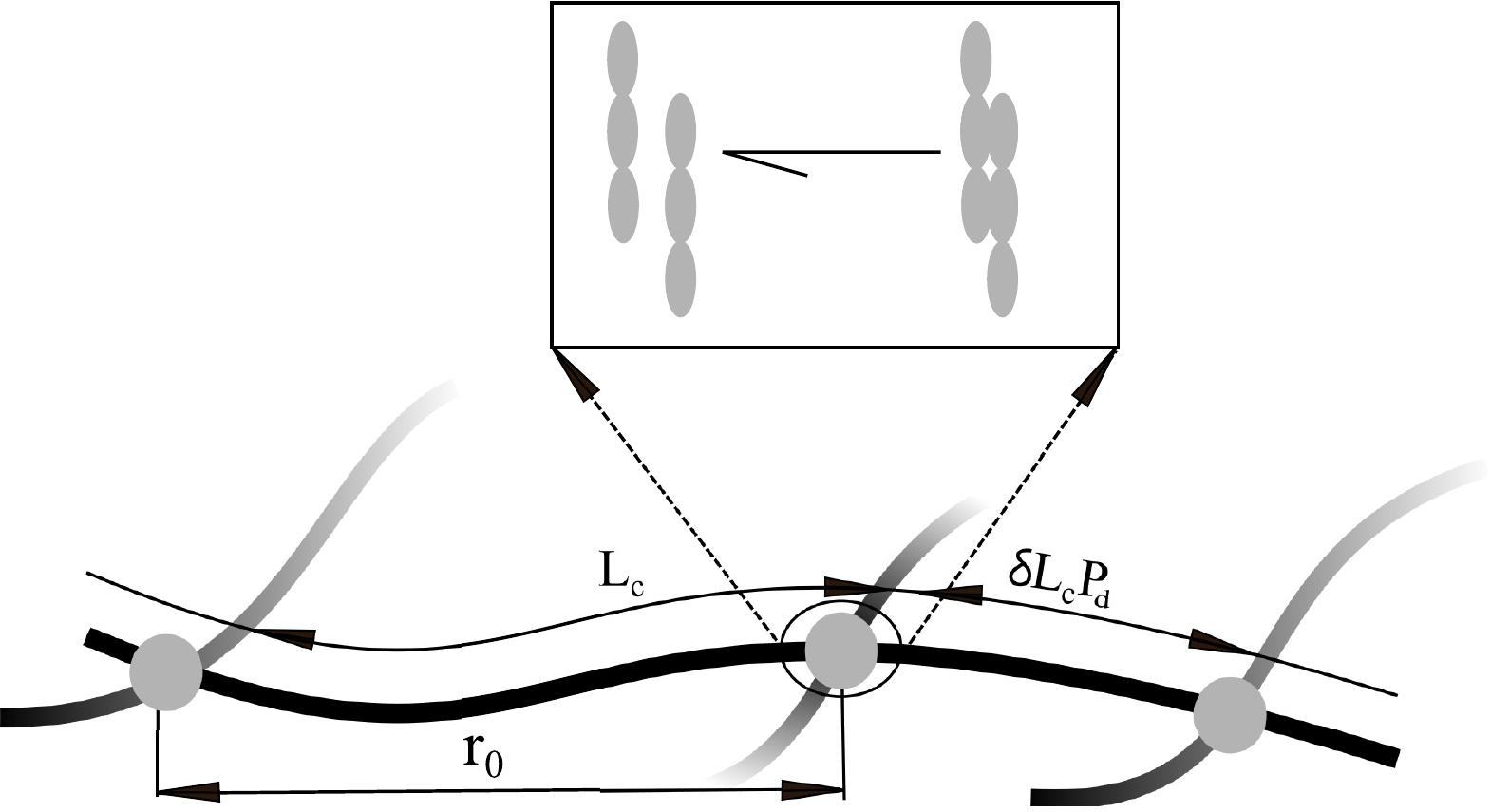} 
\caption{Semiflexible bundle structure}
\label{bundle}
\end{figure}

In order to describe the constitutive model for the mechanics of the monolayer,  we define a cytokeleton-like material model composed by an isotropic "Strain Energy Function" (SEF) which describes the role of the membrane, the intermediate filaments, and micro-tubules, and a second SEF describing the orientable F-actin network characterised by the structure tensor.
The mechanical behaviour of the F-actin cross-linked network is based on the wormlike chain model for semiflexible filaments. This model, proposed by Macintosh~\cite{Mackintosh1995}, and whose homogenised form was derived by Palmer and Boyce~\cite{Palmer2008}, is defined in terms of four physical parameters related to the network architecture and deformation (see Figure~\ref{bundle}): i) the contour length, $L_{c}$; ii) the persistence length, $l_{p}$; iii) the end-to-end length at zero force, $r_{0}$; and iv) the stretching from the condition of zero force, $\lambda$. The ability of the F-actin network to remodel its structure is introduced in the model through a dependence of the remodelling variables on the physical parameters of the network. The contour length, $L_{c}$, is assumed to be dependent on the cross-linkers density. In brief, if the cross-links unbinding probability, $P_{ub}$, increases, $L_{c}$ increases as well, i.e., $L_{c}=f(P_{ub})$.

In addition, the action of myosins in the formation of new cross-links contribute to induce pre-stress in the bundle structure. In terms of the physical parameters of the worm like chain model, $r_{0}$ is dependent on the  active action of the molecular motors, i.e., $r_{0}=f(\Delta\bar{\mu})$. Hence, we arrive to a mathematical description for the F-actin network elasticity in which the material parameters are modulated by stochastic variables associated with the chemical kinetics of remodelling
\begin{equation}
\Psi(\mathbf{C},r_{0},l_{p},L_{c})=\Psi_{iso}(\mathbf{\bar{C}})+\Psi_{wlc}(\mathbf{\bar{C}},r_{0},l_{p},L_{c})+U(J)\label{eq:W}
\end{equation}
where
\begin{equation}
\begin{split}\Psi_{wlc} & =\frac{nk_{B}T}{l_{p}}\left[\frac{L_{c}}{4\left(1-\displaystyle{\frac{r}{L_{c}}}\right)^{2}}-l_{p}\log\frac{L_{c}^{2}-2l_{p}L_{c}+2l_{p}r}{r-L_{c}}\right],\end{split}
\label{eq_3}
\end{equation}

Where $r=[r_{0}+\delta r_{0}\Delta\bar{\mu}]\lambda$.The parameter ${r_{0}}$ represents the end-to-end distance corresponding to the response of the network at zero force ,$r_0=r_{\sigma=0}=L_{c}\left(1-\frac{L_{c}}{6l_{p}}\right)$~\cite{Palmer2008,lopez2017}; and
$\delta r_{0}\Delta\bar{\mu}$ is the contribution associated with the contraction induced by the actin-myosin molecular motors. The parameter $\delta r_{0}>0$ is related to the maximum induced value of the contraction (given as a fraction of $r_{0}$ when $\Delta\bar{\mu}$ is maximum), $\Delta\bar{\mu}$ represents a normalized value of the myosin molecular motor activity. Finally, $\lambda=\sqrt{\frac{1}{3}(1-\Gamma) \bar{I}_1+ \Gamma \bar{I}_4}$ is the homogenized and macroscopic network stretch, where $\bar{I}_1$ and $\bar{I}_4$ is the first and fourth invariant of $\bar{\mathbf{C}}$.  Using standard procedures from Continuum Mechanics, the Cauchy stress,$\boldsymbol{\sigma}$, can be derived from direct differentiation
of Eq. (\ref{eq:W}) with respect to $\mathbf{C}$ 

\begin{equation}
\begin{split}\boldsymbol{\sigma} & =\frac{2}{J}\mathbf{F}\frac{\partial\Psi}{\partial\mathbf{C}}\mathbf{F}^{T}\\
 & =2C_1\mathbf{b}+\frac{nk_{B}T}{3l_{p}}\frac{r_{0}}{\lambda}\left[\frac{1}{4\left(1-\displaystyle{\frac{\lambda r_{0}}{L_{c}}}\right)^{2}}\right]\left[\frac{\displaystyle{\frac{L_{c}}{l_{p}}}-6\left(1-\frac{\lambda r_{0}}{L_{c}}\right)}{\displaystyle{\frac{L_{c}}{l_{p}}}-2\left(1-\frac{\lambda r_{0}}{L_{c}}\right)}\right]\mathbf{h}+p\mathbf{I},
\end{split}
\label{eq_5-1}
\end{equation}
where the push-forward of the structure tensor $\mathbf{h=FHF^{T}}$ is

\begin{equation}
\mathbf{h}=\frac{1}{3}\left(1- \Gamma \right)\mathbf{b}+ \Gamma \mathbf{\bar{a}_{0}}\otimes\mathbf{\bar{a}_{0}}\label{eq:h}
\end{equation}
and $\mathbf{\overline{a}_{0}}=\mathbf{Fa_{0}}$. $p=dU/dJ$ is a constitutive relation for the dilatational part of $\boldsymbol{\sigma}$.

%\subsubsection*{Fluidisation model}

Next, in order to include in the constitutive model the inelastic effects observed during the displacement of the flock of cells over the convergent channel, we define a phenomenological function for the probability of bundle binding, $P_b$, where the bundle deformation will be the driving force. We follow a similar approach as the developed by Lopez-Menendez et al. \cite{lopez2016} or Rodriguez et al.~\cite{Rodriguez2006}. The homogenized bundle deformation is defined as:
\begin{equation}
\varepsilon =\sqrt{\frac{1}{3}(1-\Gamma) \bar{I_1}+\Gamma \bar{I_4}}-1
\end{equation}

\begin{figure}[htt]
\centering
\includegraphics[width=10cm]{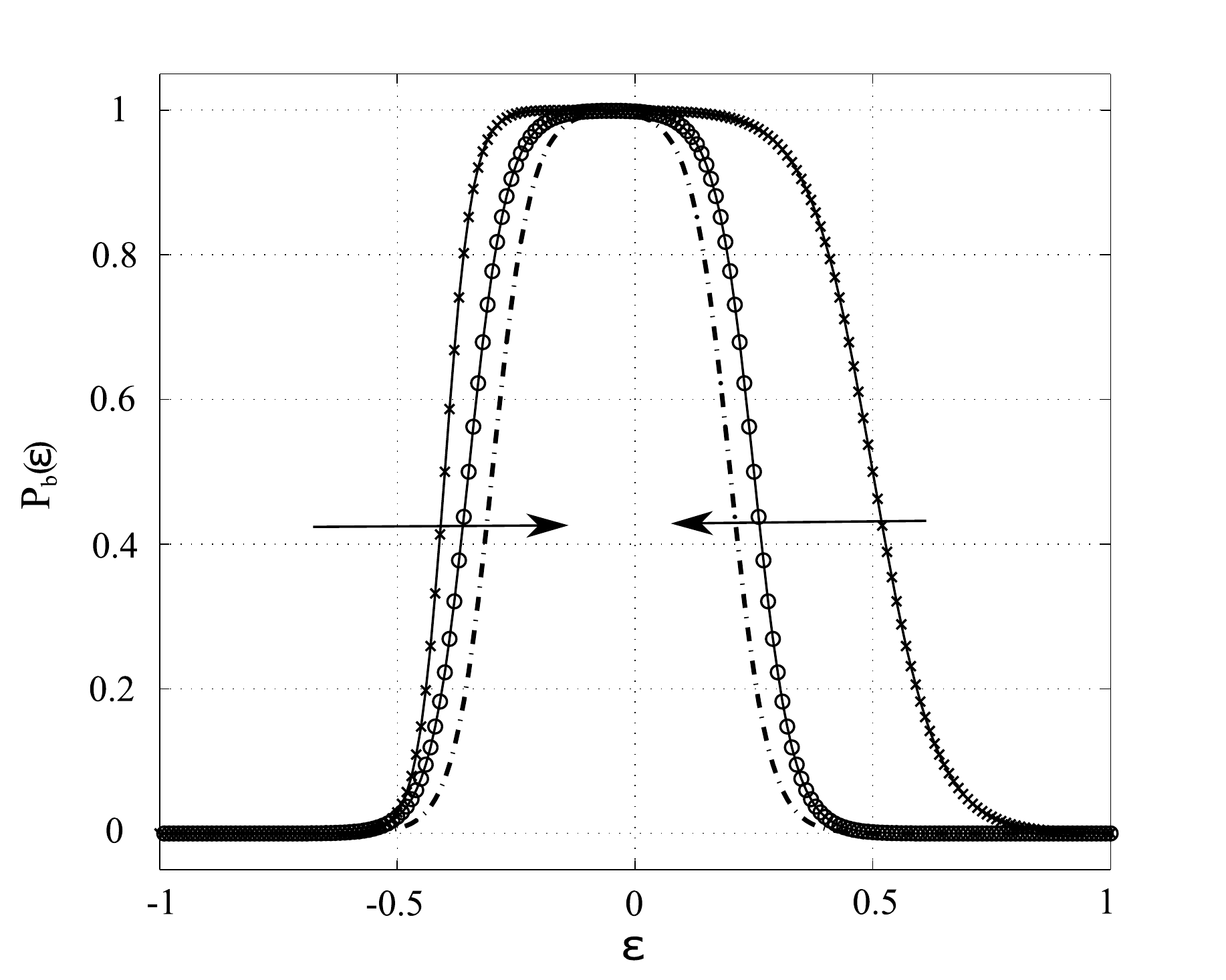} 
\caption{The phenomenological funtion $P_b$ describes the Probability of binding into the bundle when the structure is working under traction and under compression, and as can be observed this function could be asymmetrical. The arrows show the direction where the damage decreases.}\label{Pb}
\end{figure}

The functional form for $P_b(\varepsilon)$ is defined as 
\begin{equation}\label{eq_pb}
P_b(\varepsilon) =\left\{ \frac { 1 }{ 1+\exp\left[ -{ \kappa  }_{ t }(\varepsilon_{ t} -\varepsilon ) \right]  }  \right\} \left\{ \frac { 1 }{ 1+\exp\left[ { \kappa  }_{ c }(\varepsilon  +\varepsilon _{ c }) \right]  }  \right\}. 
\end{equation}
Note that the probability of binding as defined in Eq.~\ref{eq_pb} consider that the unbinding may occurs under either macroscopic traction or compression. However, as shown in Figure \ref{Pb}, the function can be non-symmetric with respect to $\varepsilon$.

The inelastic effects are introduced in the SEF proposed above by allowing the contour length $L_c$ to be the function of the probability of binding, $P_b$ 
\begin{equation}\label{eq_Lc}
L_c(\varepsilon )=L_c^{max}-\delta Lc P_b(\varepsilon ),
\end{equation}
where $L_c^{max}$ represents the maximum length of the bundle, and $\delta Lc$ is the change in the contour length due to the density of the cross-links in binding state. According to Eq.~\ref{eq_Lc}, as $P_b(\varepsilon )$ decreases, the contour length increase and the bundle stiffness decreases, leading to a local softening of the flock.

\section*{Numerical Simulation}

\begin{figure}[ht]
\centering
\includegraphics[width=12cm]{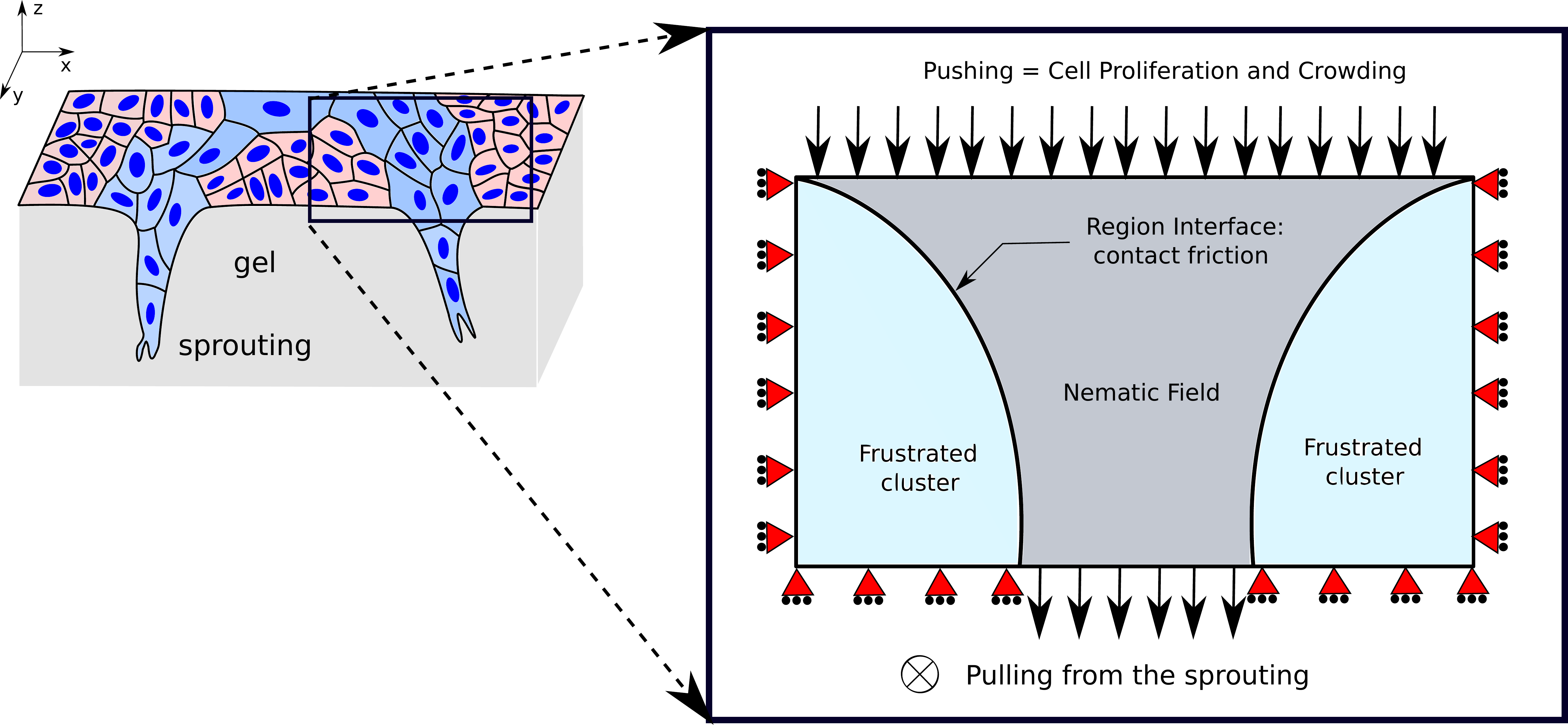}
\caption{Simplified geometry for the gate model to mimic the transition of the flock of cells from the endothelial monolayer towards the capillary-like structure}\label{domain}
\end{figure}

In this section we describe the minimal constitutive model for a physical description of a flock of cells with the ability to flow from the endothelial monolayer to the sprouting-like structure. We define a simplified region of the flock adjacent to the sprouting, formed by a convergent channel, described by a continuum of cells in a dominant unjammed phase. There are two regions adjacent to the convergent channel defined as a region of the monolayer with high density. These regions are modelled initially, for simplicity, as a rigid solid with all the displacements restricted, as can be observed in the Figure \ref{domain}.

%\begin{figure}[h]
%\centering
%\includegraphics[width=12cm]{Figures/domain.pdf}
%\caption{Simplified geometry for the gate model}\label{Fig4-1}
%\end{figure}

On top of the convergent channel we model the effects of the internal stress induced by proliferation and crowding, as an imposed displacement with an associated time history. On the bottom of the convergent channel we model the pulling effect of the capillary-like structure over the monolayer and with a traction strain with an associated time history which represents the growth of the structure. The combined action of these mechanical loads defines a field of stress and strain which induces the remodelling effects of the material, such as damage into the channel and reorientation in the direction of the sprouting.

The constitutive model presented in the previous sections has been implemented in the general purpose finite element program ABAQUS through a ``user subroutine'' (UMAT). The finite element formulation is based on the weak form of the momentum balance, the solution is carried incrementally, and the discretised nonlinear system of equations are solved using Newton's method with consistent linearisation. A mixed formulation is used to properly treat the material incompressibility. In the calculations, the integration of the cluster SEF has been performed by means of an adaptive Gauss quadrature algorithm at each Gauss point in the finite element formulation. In the remaining of the section, the first steps in the simulation of the displacement of the flock of cells are presented. The numerical solution of a boundary value problem corresponds to the idealised convergent channel as the domain proposed in the Figure~\ref{domain}. The purpose of the simulation is to demonstrate the previous described mechanism of the displacement of a flock of cells induced by the fluidisation of the a monolayer. The load imposed at the top is defined by a displacement condition, associated with the proliferation and crowding effects. At the bottom, a time dependent load associated with the pulling effect related to the growth of the capillary-like structure. The constants parameters associated with the variables of the model are presented in the Table~\ref{tabR1}.

\begin{table}[ht]
\centering
\begin{tabular}{lll}
\hline\hline 
$1+\epsilon$ & Bundle pre-strain & $1.002$\tabularnewline
$nk_BT$& Constant $\propto$ thermal energy & $2.74$ [$Pa$]\tabularnewline
$l_{p}$ & Persistence length & $10$ [$\mu$m]\tabularnewline
$L_{c}^{max}$ & Contour length & $50$ [$\mu$m]\tabularnewline
$\delta L_{c}$ & $\delta$ contour length & $40$ [$\mu$m]\tabularnewline
$\varepsilon_{t}$ & Characteristic stretch irrev. cross-linkers & $0.1$\tabularnewline
$\varepsilon_{c}$ & Characteristic stretch irrev. cross-linkers & $-0.1$\tabularnewline
$\kappa_{t}$ & Nondimensional cross-linkers stiffness & $15$\tabularnewline
$\kappa_{c}$ & Nondimensional cross-linkers stiffness & $15$\tabularnewline
$\Gamma$ & anisotropy parameter & $0.1$\tabularnewline
$C_1$ & Neo-Hookean constant & $0.5$[$Pa$]\tabularnewline
$\tau_{w}$&Average reorientation time &$100$ [$min$] \tabularnewline
\hline
\end{tabular}
\caption{Model parameters for the simulation of the extrusion}\label{tabR1}
\label{table1}
\end{table}

\section*{Results}

%\paragraph{PRINCIPAL STRESS}

\begin{figure}[h!]
\centering
\includegraphics[scale=0.45]{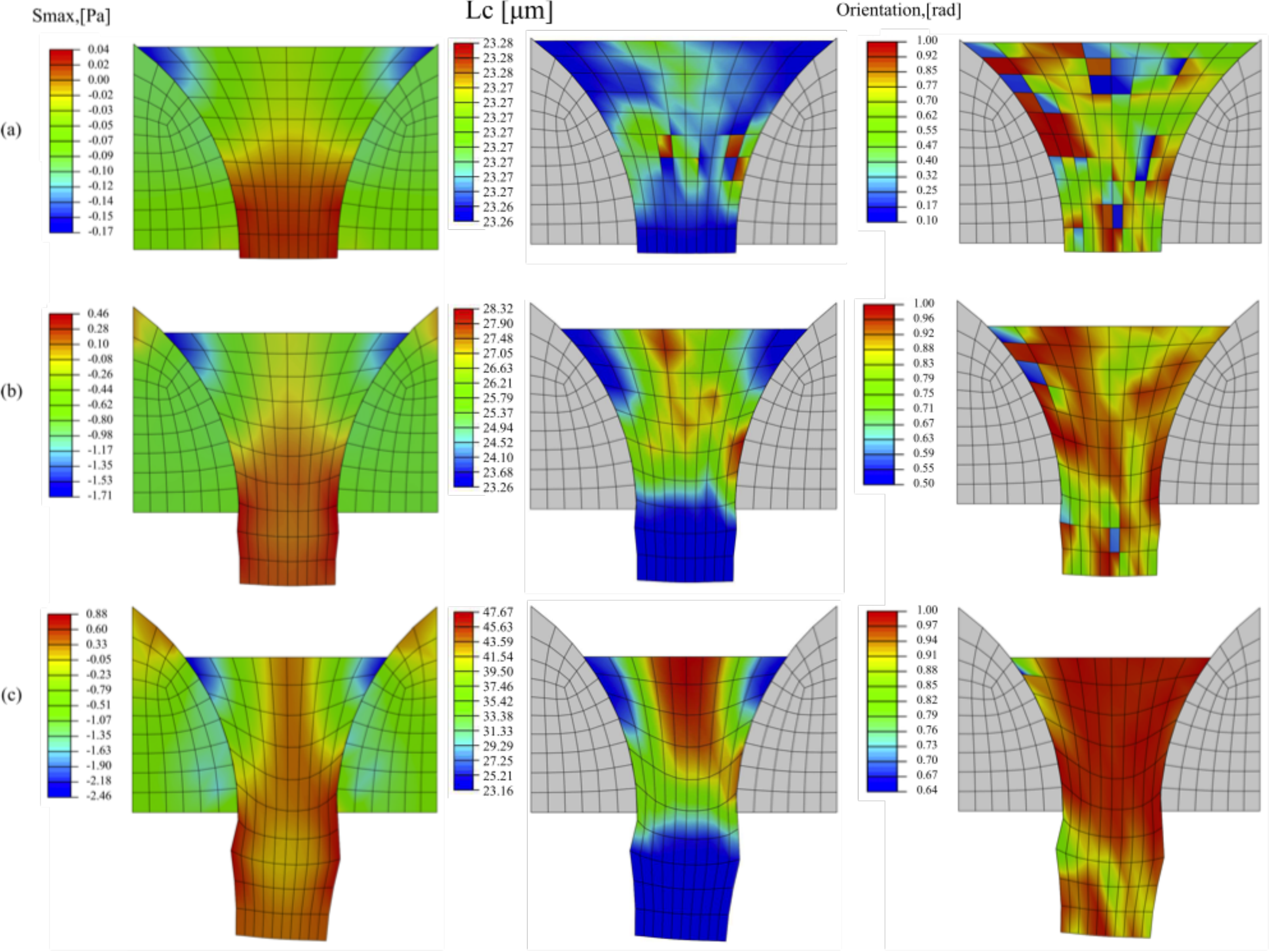}
\caption{Evaluation of the constitutive model on the simplified geometry to mimic the transition of the flock of cells from the endothelial monolayer towards the capillary-like structure. Evolution of the maximum principal stress, the orientation vector $\mathbf{a}_0$ and the damage internal variable  $L_c$  for different observed times, for (a) $t=5 min$ (b) $t=50 min$ and (c) $t=100 min$ 
 \label{flow}}
\end{figure}

The results of the numerical simulation are shown in the Figure~\ref{flow} where the first column describes the time lapsed evolution of the principal stress through the convergent channel. It can be clearly identified the increment in the stress at the bottom region due to the pulling effect introduced by the perturbation into the monolayer induced by the capillary-like structure. When the flock of cells advances, compressive stress is observed at the sides of the top layer of the moving part. However, large increments in the stress during the extrusion process are not observed due to the presence of a damage model which contributes to reduce the stiffness of the pack of cells that is passing through the sink-hole. Note that the maximum stress is found in the center of the channel rather than on the sides next to the channel like structure. Once the cells pass through the sprouting, they try to get back to their original state due to the neo-Hookean terms of the energy function.

%\paragraph{ORIENTATION}

%\paragraph{DAMAGE}
Here we choose to reflect the effects of damage in a phenomenological manner over the value of the bundle contour length $L_c$, which is associated with the cytoskeleton's F-actin network mesh-size. As we explained before, within the model of semiflexible $wlc$, the stiffness and the contour length show an inversely proportional relation, and clearly an increment in the pore-size of the network gives us an idea of the damage of the structure. The Figure~\ref{flow}, in the second column,  describes the spatial distribution of the $L_c$ for different times of the loading history. The figure shows that $L_c$ increases as the flock advances through the channel. The maximum value of the variable is observed in the central region of the flock, implying a larger amount of damage, i.e., cross linking rupture. In terms of the $wlc$ model, it indicates a larger reduction in the network stiffness in the central region of the channel with respect to the sides, which makes easier to get through the channel. 

The Figure~\ref{flow}, in the third column, depicts three representative stages of the cells orientation in the monolayer during the loading history. The top figure shows the initial state. The pattern of colours indicates the initial orientation of the homogenized flock of cells, in the direction $\mathbf{a}_0$, that have been assigned randomly at each integration point. As time advances, the homogenised flock aligns in the direction of the polarised field defined by the sprouting and the strain over the monolayer. When the cells became part of the capillary-like structure, they keep the defined oriented direction and thus preserve their shape.

%%% Cell division
%This effect was not considered into the material model, in future works can be introduced. Here was considered as whole inside the displacement field applied on the top of the convergent channel.

\section*{Discussions \& Conclusion}

To generally sum up, we notice that previous works mainly define the angiogenic process from the sprouting point of view. To challenge this idea, we put forward the hypothesis that the angiogenesis, via the action of the VEGF , can cause the collective unjamming of the endothelial cells, which in combination with the internal relaxation of the tissue, encodes the displacement of the flock of cells from the monolayer. 

Thus we arrive to a new mechanical picture for the angiogenic process analysing the unjamming of the monolayer made by angiogenic growth factors. First, we conclude that the role of the \textit{cell shape index}  as an estimator of the monolayer fluidisation during angiogenesis, establishes a simple structural index that quantifies the proximity of the layer to an unjamming transition. This suggests that, perhaps, molecular mechanisms can be investigated and categorised by how they affect the jamming (Figure 1). Once we proved quantitatively the effect of jamming - unjamming transition induced by the VEGF, we measured the \textit{density field} in the drug treated monolayer and we  found the interaction between fluidisation and capillary-like structure. 

Here , we provide a novel mechanical picture to think on angiogenesis, by looking in-vitro endothelial monolayer from the top and studying the alterations over the cells density. Strikingly, we found that  a similarity between the morphogenic process into the monolayer during angiogenesis and the  'evacuation process' can be established. Also, we identified some regions in vicinity of the sprouting,  where the cell density increases. This is associated with the bottleneck by which the cells cannot flow and are frustrated. In addition ,  as is observed during the evacuations process, some cells are able to flow across the sprouting forming a queue,  or  a nematic field to get out, where the cells re-orient themselves towards the sprouting. In summary, the combination among regions of high and low density of cells, defines what we call the convergent channel by which a flock of cells can flow from the monolayer to the capillary-like structure (Figure 2).

These observations suggest some questions associated to the role of the cell proliferation and the interaction with the angiogenesis. On this regard, previous works indicate that when the VEGF is activated over the endothelial monolayer, cells undergo division through an independent process of endothelial growth factor (EGF) \cite{Semino2006}.  As we report here, the heterogeneities on the cell density are proportional with the level of strain. We suspect that the proliferative action can be coupled with this effect:  Gudipaty et al. proved it on an epithelial monolayer under stretch, where the cell division is promoted via the activation of a pathway related to the  Piezo-1 stretch-channel \cite{gudipaty2017}. It suggests an interesting positive feedback by which, more stretch over the monolayer, produced by the capillary-like structure, enhances cells proliferation inside the tissue.

Another important aspect of these observations is the relation between orientation and division. When cells are inside the nematic field pointing in the direction of the sprouting, they also orient  themselves in the same direction as the spindle. This coupling was previously shown on single cells by Fink et al. \cite{fink2011}, and tissues by Campinho et al. \cite{campinho2013}. The cell division represents a source of internal stress, as they must rearrange to make room for the new cells \cite{Ranft2010}. In addition, this stress is anisotropic , occurring over a preferred cell axis and, potentially,  it enhances the flow of cell towards the capillary-like structure.  Nevertheless, future works will be necessary to better understand how cell division can affect to the evolution of the angiogenic process. In our opinion, at least, two scenarios result plausible. On the one hand, the cell proliferation can enhance the bottle-neck and thus the cells became frustrated to flow from the monolayer. On the other hand, if the resultant configuration defines the convergent channel, as we mentioned before, a cooperative process where more cells divide and flow can be defined. 

Taking together our observations and novel mechanical insight, we conclude that, in order to obtain a better understanding on the angiogenesis, we should take a perspective, from the collective migration of cells \cite{ladoux2017}, where the flock of cells can flow from the monolayer towards the capillary in a coordinated manner. Future works correlating experimental observations will reveal what is the relation between the mechanical disturbance of the cell monolayer due to the sprouting and the internal pressure due to proliferation and crowding. They will also reveal the velocity of reorientations, and the concentration of drugs, as the VEGF, associated with the remodelling of the cytoskeleton structure. These speculations are far to be closed. Nevertheless, we think that new ways will emerge if we are capable to understand the natural mechanism between the endothelial monolayer and the capillary-like structure. Likewise,  it remains to be determined if these described effects can provide better insights to help the development of techniques to block them,  as in the case of the oncogenic drugs.

\section*{Acknowledgments}
We thanks to Prof. Carlos Semino from Universitat Ramon Llull, for sharing the original images for the analysis performed in this work. Also to the Prof. Jos\'e Felix Rodr\'iguez, from Milano Polytechnic for his help in the simulation and the fruitful discussion. To the Prof. Jos\'e Manuel Garcia Aznar from University of Zaragoza for his interest in the early steps of this project. To the Prof. Jacques Prost from Institute Curie for his valuable feedback. 

\section*{Conflicts of interest}
The authors declare that they have no affiliations with or involvement in any organisation or entity with any financial interest,  or non-financial interest in the subject matter or materials discussed in this manuscript.

\section*{Apendix 1}
\subsection*{Stress and Elasticity tensors}

For a hyperelastic material with a SEF, $\Psi$, defined as in
Eq.~\ref{SEF_anisotropic_uncoup}, the second Piola-Kirchhoff stress can be
written as

\begin{equation}\label{piola_str1}
\mathbf{S}=2\frac{\partial\Psi}{\partial\mathbf{C}}=
pJ\textbf{C}^{-1}+2J^{-2/3}\mathrm{DEV}\left[\frac{\partial
\bar{\Psi}}{\partial \bar{\textbf{C}}}\right],
\end{equation}
where $p=U'(J)$, is the hydrostatic pressure, and $DEV[\cdot]$ is the
deviatoric projection operator in the material description
\begin{equation}
\mathrm{DEV}[\cdot]\equiv[\cdot]-\frac{1}{3}([\cdot]:\bar{\textbf{C}})\bar{\textbf{C}}^{-1}.
\end{equation}
The Cauchy stress tensor is found by the weighted pushed forward of Eq.~\ref{piola_str1}
\begin{equation}\label{cauchy_str1}
\boldsymbol{\sigma}=J^{-1}\mathbf{FSF}^T=
p\textbf{1}+2J^{-1}\mathrm{dev}\left[\bar{\textbf{F}}\frac{\partial
\bar{\Psi}}{\partial \bar{\textbf{C}}}\bar{\textbf{F}}^T\right],
\end{equation}
where
\begin{equation}
\mathrm{dev}[\cdot]\equiv[\cdot]-\frac{1}{3}([\cdot]:\textbf{1})\textbf{1}.
\end{equation}
The material version of the elasticity tensor is defined as
\begin{equation}\label{C_mat}
\mathbb{C}:=2\frac{\partial\mathbf{S}}{\partial\mathbf{C}}=
4\frac{\partial^2\Psi}{\partial\mathbf{C}\partial\mathbf{C}}.
\end{equation}
For the uncoupled SEF Eq.~\ref{SEF_anisotropic_uncoup}, it is given by
\begin{equation}\label{C_mat2}
\begin{array}{rcl}
\mathbb{C}&=&pJ(\textbf{C}^{-1}\otimes\textbf{C}^{-1}+2\mathbb{I}_{C^{-1}})-
\frac{4}{3}J^{4/3}\left(\frac{\partial \bar{\Psi}}{\partial
\bar{\textbf{C}}}\otimes\textbf{C}^{-1}+\textbf{C}^{-1}\otimes\frac{\partial
\bar{\Psi}}{\partial \bar{\textbf{C}}}\right)\\
&&+\frac{4}{3}\left(\frac{1}{3}\bar{\textbf{C}}^{-1}\otimes\bar{\textbf{C}}^{-1}-
\mathbb{I}_{\bar{C}^{-1}}\right)\left(\frac{\partial \bar{\Psi}}{\partial
\bar{\textbf{C}}}:\bar{\textbf{C}}\right)+\frac{4}{3}J^{4/3}\bar{\mathbb{C}}_{\bar{\Psi}},
\end{array}
\end{equation}
where
\begin{equation}
(\mathbb{I}_{C^{-1}})_{IJKL}=\frac{1}{2}(C^{-1}_{IK}C^{-1}_{JL}+C^{-1}_{IL}C^{-1}_{JK})
\end{equation}
and
\begin{equation}
\begin{array}{rcl}
\bar{\mathbb{C}}_{\bar{\Psi}}&=&3\frac{\partial^2 \bar{\Psi}}{\partial
\bar{\textbf{C}}\partial \bar{\textbf{C}}}-\left[ \left(\frac{\partial^2
\bar{\Psi}}{\partial \bar{\textbf{C}}\partial
\bar{\textbf{C}}}:\bar{\textbf{C}}\right)\otimes\bar{\textbf{C}}^{-1}
+\bar{\textbf{C}}^{-1}\otimes\left(\frac{\partial^2 \bar{\Psi}}{\partial
\bar{\textbf{C}}\partial
\bar{\textbf{C}}}:\bar{\textbf{C}}\right)\right] \\
&&+\frac{1}{3}\left(\bar{\textbf{C}}:\frac{\partial^2 \bar{\Psi}}{\partial
\bar{\textbf{C}}\partial \bar{\textbf{C}}}:\bar{\textbf{C}}
\right)\bar{\textbf{C}}^{-1}\otimes\bar{\textbf{C}}^{-1}.
\end{array}
\end{equation}
A similar weighted push-forward operation of the elasticity tensor leads to its
spatial counterpart, $\scriptstyle{\mathbb{C}}$,\cite{Spencer1980}
\begin{equation}\label{c_espatial}
\begin{array}{rcl}
{\scriptstyle\mathbb{C}}&=&p(\textbf{1}\otimes\textbf{1}-2\mathbb{I})-\frac{2}{3}(
\mathrm{dev}\boldsymbol{\sigma}\otimes\textbf{1}+\textbf{1}\otimes\mathrm{dev}\boldsymbol{\sigma})
\\
&&+\frac{4}{3J}\left(\frac{\partial \bar{\Psi}}{\partial
\bar{\textbf{C}}}:\bar{\textbf{C}}\right)\left[\mathbb{I}-\frac{1}{3}(\textbf{1}\otimes\textbf{1})\right]+
{\scriptstyle\mathbb{C}_{\bar{\Psi}}},
\end{array}
\end{equation}
where ${\scriptstyle\mathbb{C}_{\bar{\Psi}}}$ is the weighted pushed forward of
$\mathbb{C}_{\bar{\Psi}}$, and $\mathbb{I}$ the fourth order identity tensor.
These expressions will be used in the finite element implementation described
in subsequent sections.

\section*{Apendix 2}
\subsection*{Measurement of the density field}
The density field was measured from the positions of nuclei, using homemade ImageJ and Matlab programs whose main steps are described in what follows. First the nuclei images were z-projected, using the maximum intensity and filtrated using a low-pass FFT filter with a minimum wavelength of 5 pixels. Then the positions were retrieved using the Find Maxima function of ImageJ, with a single noise tolerance determined manually for each experiment. Next, the positions were loaded into Matlab and their number counted in a sliding window of 150 pixels with an interval of 10 pixels between each measurement center. Finally, the density field was smoothed in space and time using a box filter of dimensions 3x3 pixels and 9 frames.

%\bibliographystyle{unsrt}
%\addcontentsline{toc}{section}{\refname}\bibliography{Angio}
\end{document}